\documentclass[12pt]{article}

\begin{document}

\title{Equivalence Between Space-Time-Matter and Brane-World Theories}
\author{J. Ponce de Leon\thanks{E-mail: jponce@upracd.upr.clu.edu or jpdel@astro.uwaterloo.ca}\\ Laboratory of Theoretical Physics, Department of Physics\\ 
University of Puerto Rico, P.O. Box 23343, San Juan, \\ PR 00931, USA \\ 
Department of Physics, University of Waterloo,\\
 Waterloo, Ontario N2L 3G1, Canada.}
\date{November 2001}

\maketitle
\begin{abstract}
We study the relationship between space-time-matter (STM) and brane theories.  
These two theories look very different at first sight, and have different motivation for the introduction of a large extra dimension. However, we show that they are equivalent to each other. 
First we demonstrate that STM predicts local and non-local high-energy corrections to general relativity in $4D$, which are identical to those predicted by brane-world models. 
Secondly, we point out that in brane models the usual matter in $4D$ 
is a consequence of the dependence of five-dimensional metrics on the extra coordinate. 
If the $5D$ bulk metric is independent of the extra dimension, then the brane is void of matter. Thus, in brane theory matter and geometry are unified, which is exactly the paradigm proposed in STM.
Consequently, these two $5D$ theories share the same concepts and predict the same physics. This is important not only from a theoretical point of view, but also in practice. We propose to use a combination of both methods to alleviate the difficult task of finding solutions on the brane. We show an explicit example that illustrate the feasibility of our proposal. 

 \end{abstract}

PACS: 04.50.+h; 04.20.Cv 

{\em Keywords:} Kaluza-Klein Theory; Brane Theory; General Relativity

\newpage

\section{INTRODUCTION}

The basic ideas of what today is called space-time-matter  theory (STM) have been developed  by a number of people \cite{Wesson 1}-\cite{JPdeL 2}. In this theory, our four-dimensional world is embedded in a five-dimensional spacetime, which is a solution of the five-dimensional Einstein's equations in vacuum.  The extra dimension is not assumed to be compactified, which is a mayor departure from earlier multidimensional theories where the cilindricity condition was imposed. In this theory, the original motivation for assuming the existence of a large extra dimension  was to achieve the unification of matter and geometry, i.e., to obtain the properties of matter as a consequence of the extra dimensions.

Recent results in string theories suggest that gravity is indeed a multidimensional interaction, and that usual general relativity in $4D$ is the low energy limit of some more general theory. 
In these theories the matter fields are confined to our $4D$ spacetime, embedded in a $4 + d$ dimensional spacetime, while gravity fields propagate in the extra $d$ dimensions as well. 
In $5D$ a number of works model our $4D$ universe as a domain wall in a five-dimensional anti-de Sitter spacetime. In this context, the motivation  for large extra dimensions is to solve the hierarchy problem \cite{Randall1}-\cite{Randall2}. 

This has promoted an increasing interest in gravity theories formulated in spacetimes with large extra dimensions. Brane-world models are inspired by these theories. In these models our universe is a  four dimensional singular hypersurface, or ``brane", in a five-dimensional spacetime, or bulk \cite{Shiromizu}-\cite{Maartens2}.

Although STM and brane theory have different physical motivations for the introduction of a large extra dimension, they share the same working scenario. Namely, (i) they allow the bulk metrics to have non-trivial dependence of the extra dimension; (ii) the $4D$ metric is obtained by evaluating the background metric at some specific $4D$ hypersurface that we identify with our physical spacetime;
(iii) the matter fields are confined to a 3-brane in a $5D$ spacetime, which is a solution to Einstein's equations; (iv) observers are bound to the brane, unable to access the bulk. 

From a practical point of view, they share the same goals. Among them, to predict the effects of the bulk geometry on the brane geometry and dynamics. 

Despite of these common grounds, both theories remain as separated, unrelated, entities.
The aim of this work is to remedy this situation. Our first goal is, therefore, to show that both theories are equivalent. We will show that STM  includes the so-called local high-energy corrections, and non-local Weyl corrections typical of brane-world scenarios. Also that the matter in the brane is purely geometric in nature.

The difference in the motivation for large extra dimensions is reflected in the techniques adopted by authors in each theory. In STM the authors start from the geometry of the bulk and construct the physics in $4D$ \cite{Wesson book}. In brane models, the opposite point of view is taken. Namely, the effective equations in $4D$ are solved in the brane for some matter distribution on it. Then the solution is matched to  some appropriate bulk metric that satisfies Israel's boundary conditions \cite{Dadhich}-\cite{Cristiano}. 

The problem of finding a complete solution in brane theory is a really involved one. Therefore, 
our second goal in this work is to propose we use the formal equivalence between the two theories to alleviate this problem. Technically, what we propose is that we incorporate the physics of brane models to interpret and determine the final form of solutions in STM. We show the feasibility of this approach by means of an explicit example. 

\section{Equivalence Between STM and Brane Theory} 

In this Section we first show that STM  incorporates and predicts the same physics as brane-world models. We then show that brane models include and share with STM the same philosophy about the geometrical nature of matter in $4D$. Finally, we discuss the different points of view adopted in STM and brane-world models, regarding physics in $4D$. 

\subsection{Equations In Space-Time-Matter Theory}

The metric is taken as 
\begin{equation}
\label{5D metric restricted metric}
d{\bf {\cal S}}^{2} = g_{\mu \nu}(x^\rho,y) dx^{\mu} dx^{\nu} + \epsilon \Phi^2(x^\rho,y) dy^{2},
\end{equation}
where $\epsilon$ is $- 1$ or $+ 1$ depending on whether the extra dimension is spacelike or timelike, respectively. In what follows $\stackrel{\ast}{f} = \partial f/\partial y$,
and the covariant derivatives are taking with respect to $g_{\mu\nu}$. For the signature of the spacetime and definition of tensor quantities we follow Landau and Lifshitz \cite{Landau and Lifshitz}.

The field equations in STM theory are the Einstein equations in vacuum, viz., $^{(5)}R_{AB} = 0$. The $4+1$ splitting of these equations provides a definition for an effective energy-momentum tensor \cite{Wesson and JPdeL}
\begin {equation}
\label{EMT in STM}
^{(4)}G_{\alpha\beta} = \frac{\Phi_{\alpha;\beta}}{\Phi} - \frac{\epsilon}{2\Phi^2}\left[\frac{\stackrel{\ast}{\Phi} \stackrel{\ast}{g}_{\alpha \beta}}{\Phi} - \stackrel{\ast \ast}{g}_{\alpha \beta} + g^{\lambda\mu}\stackrel{\ast}{g}_{\alpha\lambda}\stackrel{\ast}{g}_{\beta\mu} - \frac{1}{2}g^{\mu\nu}\stackrel{\ast}{g}_{\mu\nu}\stackrel{\ast}{g}_{\alpha\beta} + \frac{1}{4}g_{\alpha\beta}\left(\stackrel{\ast}{g}^{\mu\nu}\stackrel{\ast}{g}_{\mu\nu} + (g^{\mu\nu}\stackrel{\ast}{g}_{\mu\nu})^2\right)\right],
\end{equation}
an equation governing the scalar field $\Phi$  
\begin{equation}
\label{wave equation}
\epsilon \Phi \Phi^{\mu}_{;\mu} = - \frac{1}{4}\stackrel{\ast}{g}^{\lambda\beta}\stackrel{\ast}{g}_{\lambda\beta} - \frac{1}{2}g^{\lambda\beta}\stackrel{\ast \ast}{g}_{\lambda\beta} + \frac{\stackrel{\ast}{\Phi}}{2\Phi}g^{\lambda\beta}\stackrel{\ast}{g}_{\lambda\beta},
\end{equation}
and a conservation equation, which is usually written as 
\begin{equation}
\label{conservation equation}
P^{\beta}_{\alpha;\beta} = 0
\end{equation}
where
\begin{equation}
\label{EMT in the brane}
P_{\alpha\beta} = \frac{1}{2\Phi}(\stackrel{\ast}{g}_{\alpha\beta} -g_{\alpha\beta}g^{\mu\nu}\stackrel{\ast}{g}_{\mu\nu}),
\end{equation}
All the above quantities are evaluated at the $4D$ hypersurface $y = y_{0} = Constant$, which is identified with the physical spacetime $\Sigma$. Thus $4D$ quantities depend on $x^0, x^1, x^2, x^3$ only, but not on $y$.

The above equations form the basis of STM. From a four-dimensional point of view, the empty $5D$ equations  look as the Einstein equations with (effective) matter. In the case where $g_{\mu\nu}$ is independent of $y$, equations (\ref{EMT in STM}) and (\ref{wave equation}) show  that the effective energy-momentum tensor is traceless, $T^{\mu}_{(eff) \mu} = 0$. In other words, independence of the $4D$ metric from the extra coordinate implies a radiation-like equation of state. Thus the existence of other forms of matter crucially depend on the derivatives of $g_{\mu\nu}$ with respect to the extra dimension. In this case the symmetric tensor (\ref{EMT in the brane}) is a non-trivial conserved matter quantity (\ref{conservation equation}). A detailed investigation shows that we can recover all of the equations of state commonly used in astrophysics and cosmology \cite{JPdeL 1}, \cite{JPdeL Wesson}-\cite{Billyard}.

The right-hand-side of (\ref{EMT in STM}) can be expressed in terms of geometrical quantities. For this we introduce the normal vector orthogonal to spacetime $\Sigma$. It is $n^A = {\delta^{A}_{4}}/{\Phi}$. Thus 
\begin{equation}
 n_{A}= (0, 0, 0, 0, \epsilon \Phi).
\end{equation} 
Then, the first partial derivatives of the metric with respect to $y$ can be interpreted in terms of the extrinsic curvature of $\Sigma$. Namely,
\begin{equation}
\label{extrinsic curvature}
K_{\alpha\beta} = \frac{1}{2}{\cal{L}}_{n}g_{\alpha\beta} = \frac{1}{2\Phi}\stackrel{\ast}{g}_{\alpha\beta},
\end{equation}
and $K_{A4} = 0$. We thus have $K_{\lambda}^{\lambda} = (g^{\alpha\beta}\stackrel{\ast}{g}_{\alpha\beta}/2\Phi)$, $K_{\alpha\beta}K^{\alpha\beta} = - (\stackrel{\ast}{g}^{\mu\nu}\stackrel{\ast}{g}_{\mu\nu}/4\Phi^2)$ and $K_{\mu\alpha}K^{\alpha}_{\nu} = (\stackrel{\ast}{g}_{\mu\alpha}\stackrel{\ast}{g}_{\nu\rho}g^{\rho\alpha}/4\Phi^2)$. For the second derivatives $\stackrel{\ast \ast}{g}_{\mu\nu}$, we evaluate $^{(5)}R_{\mu4\nu4}$. We obtain,
\begin{equation}
\label{STM in terms of geometry}
^{(5)}R_{\mu4\nu4} = -\epsilon \Phi \Phi_{\mu;\nu} - \frac{1}{2}\stackrel{\ast \ast}{g}_{\mu\nu} + \frac{1}{2}\stackrel{\ast}{g}_{\mu\nu}\frac{\stackrel{\ast}{\Phi}}{\Phi} + \frac{1}{4}g^{\rho\sigma}\stackrel{\ast}{g}_{\rho\mu}\stackrel{\ast}{g}_{\sigma\nu}. 
\end{equation}
Now substituting this expression  into (\ref{EMT in STM}) and using (\ref{extrinsic curvature}) we get
\begin{equation}
\label{4D Einstein Eq. in terms of K}
^{(4)}G_{\mu\nu} = \epsilon \left[ K^{\alpha}_{\alpha} K_{\mu\nu} - K_{\mu\alpha}K^{\alpha}_{\nu} + \frac{1}{2}g_{\mu\nu}\left(K_{\alpha\beta}K^{\alpha\beta} - (K^{\alpha}_{\alpha})^2\right) - E_{\mu\nu}\right],
\end{equation}
where
\begin{equation}
E_{\mu\nu} = \frac{^{(5)}R_{\mu4\nu4}}{\Phi^2}.
\end{equation}

Equation (\ref{4D Einstein Eq. in terms of K}) suggests to us that matter may be purely geometrical in origin. This interpretation is the backbone of STM.

Notice that, as a consequence of the field equations, $^{(5)}R_{AB} = 0$, in STM theory the Riemann tensor and the Weyl tensor become identical to each other, viz, $^{(5)}R_{ABCD} =  ^{(5)}C_{ABCD}$. Consequently,  $E_{\mu\nu} =  ^{(5)}C_{A\mu B\nu}n^A n^B$. Now, by virtue of (\ref{wave equation}), $E_{\mu\nu}$ is traceless, as one expected
\begin{equation}
E_{\alpha}^{\alpha} = 0.
\end{equation}

Let us now consider the tensor quantity (\ref{EMT in the brane}). In terms of the extrinsic curvature it becomes  
\begin{equation}
\label{P in terms of the extrinsic curvature}
P_{\alpha\beta} = K_{\alpha\beta} - g_{\alpha\beta}K^{\mu}_{\mu}.
\end{equation}
We now decompose it as
\begin{equation}
\label{decomposition of P}
P_{\mu\nu} = - \frac{k_{(5)}^2}{2}(- \lambda g_{\mu\nu} + T_{\mu\nu}),
\end{equation}
where $k_{(5)}^2$ is a constant introduced for dimensional reasons. Although this decomposition can be ambiguous, we will see that in brane-world models $\lambda$ is the tension of the brane in five-dimensions,  and $T_{\mu\nu}$ the energy-momentum tensor of the matter in the brane.
From the above equations we get
\begin{equation}
\label{K in terms of matter in the brane}
K_{\mu\nu} = - \frac{1}{2}k_{(5)}^2 \left(T_{\mu\nu} - \frac{1}{3}g_{\mu\nu}(T - \lambda)\right).
\end{equation}
Substituting (\ref{K in terms of matter in the brane}) into (\ref{4D Einstein Eq. in terms of K}), we obtain the STM effective energy-momentum tensor as
\begin{equation}
\label{EMT in brane theory}
^{(4)}G_{\mu\nu} = - {\Lambda}_{(4)}g_{\mu\nu} + 8\pi G_{N}T_{\mu\nu} +\epsilon k_{(5)}^4\Pi_{\mu\nu} - \epsilon E_{\mu\nu},
\end{equation}
where
\begin{equation}
\label{definition of lambda}
\Lambda_{(4)} = \epsilon \frac{\lambda^2 k_{(5)}^4}{12},
\end{equation}
\begin{equation}
8 \pi G_{N} = \epsilon \frac{\lambda k_{(5)}^4}{6},
\end{equation}
and
\begin{equation}
\label{quadratic corrections}
\Pi_{\mu\nu} = - \frac{1}{4} T_{\mu\alpha}T^{\alpha}_{\nu} + \frac{1}{12}T T_{\mu\nu} + \frac{1}{8}g_{\mu\nu}T_{\alpha\beta}T^{\alpha\beta} - \frac{1}{24}g_{\mu\nu}T^2.
\end{equation}
From (\ref{conservation equation}) and (\ref{decomposition of P}) it follows that
\begin{equation}
\label{conservation for T}
T^{\mu}_{\nu;\mu} = 0.
\end{equation}
Also, $^{(4)}G^{\mu}_{\nu;\mu} = 0$ implies
\begin{equation}
\label{sources for Weyl tensor}
E^{\mu}_{\nu;\mu} = k_{(5)}^4 \Pi^{\mu}_{\nu;\mu}.
\end{equation}

Equations (\ref{EMT in brane theory}), (\ref{conservation for T}) ad (\ref{sources for Weyl tensor}) are equivalent to the original set (\ref{EMT in STM}), (\ref{wave equation}) and (\ref{conservation equation}). We stress the fact that the above equations contain neither reference, nor any particular assumption, specific to the brane-world scenario. 

\subsection{STM as Generating $5D$ Space for Brane-World Models} 

It should be noted that (\ref{EMT in brane theory}), (\ref{conservation for T}) and (\ref{sources for Weyl tensor}) look exactly as the equations for gravity in brane-world models, with $\lambda$ and $T_{\mu\nu}$ as the vacuum energy and energy-momentum tensor, respectively in the brane \cite{Maartens2}. Equation (\ref{EMT in brane theory}) includes the  five-dimensional corrections to the field equations in $4D$ general relativity. These are the local quadratic corrections given by $\Pi_{\mu\nu}$ and non-local Weyl corrections from the free gravitational field in the bulk given by $E_{\mu\nu}$. 

The usual scenario in brane-world models is that matter fields are confined to a singular $3$-brane. Therefore, to proceed with our discussion we need to construct such a brane from STM. For convenience the coordinate $y$ is chosen such that the hypersurface $\Sigma: y = 0$ coincides with the brane, which is assumed to be ${\bf Z}_2$ symmetric in the bulk background  \cite{Randall1}-\cite{Maartens2}. The brane is obtained by a simple ``copy and paste" procedure. Namely, we cut the generating $5D$ spacetime, with metric $g_{AB}$, in two pieces along $\Sigma$, then copy the region $y \geq 0$ and paste it in the region $y \leq 0$. The result is a singular hypersurface in a ${\bf Z}_2$ symmetric universe with metric
\begin{equation}
\label{5D bulk metric }
d{\bf {\cal S}}^{2} = g^{bulk}_{\mu\nu}(x^\rho,y) dx^{\mu} dx^{\nu} + \epsilon dy^{2},
\end{equation}
where
\begin{eqnarray}
\label{Z2 symmetry}
g^{bulk}_{\alpha\beta} = g_{\alpha\beta}(x^\mu, + y)\;\;\;\;\  for\;\;\;\; y \geq 0,\nonumber \\
g^{bulk}_{\alpha\beta} = g_{\alpha\beta}(x^\mu, - y)\;\;\;\;\  for\;\;\;\; y \leq 0.
\end{eqnarray}
Therefore, the field equations in the bulk exhibit a delta-like singularity, viz., $^{(5)}G^{bulk}_{AB} = k_{(5)}^2[T^{bulk}_{AB} + \delta(y)S_{AB}]$, where $S_{AB}n^A = 0$ represents the total energy-momentum in the brane.

Israel's boundary conditions imply
\begin{equation}
K_{\mu\nu\mid \Sigma^{+}} - K_{\mu\nu\mid \Sigma^{-}} = - k_{(5)}^2\left(S_{\mu\nu} - \frac{1}{3}g_{\mu\nu}S\right).
\end{equation} 
This equation, together with the imposition of the ${\bf Z}_{2}$-symmetry, keeping the brane fixed, leads to 
\begin{equation}
 K_{\mu\nu\mid \Sigma^{+}} = - K_{\mu\nu\mid \Sigma^{-}} = - \frac{1}{2} k_{(5)}^2\left(S_{\mu\nu} - \frac{1}{3}g_{\mu\nu}S\right).
\end{equation}
Then from (\ref{P in terms of the extrinsic curvature}) we get
\begin{equation}
\label{P in terms of S}
P_{\mu\nu} = - \frac{1}{2}k_{(5)}^2 S_{\mu\nu}.
\end{equation}
Consequently, from (\ref{decomposition of P}) and (\ref{P in terms of S}) we find
\begin{equation}
\label{decomposing the brane matter tensor}
S_{\mu\nu} = - \lambda g_{\mu\nu} + T_{\mu\nu},
\end{equation}
which is the usual separation for the energy-momentum tensor in brane models, where $\lambda$ and $T_{\mu\nu}$ are the vacuum energy and the energy-momentum tensor, respectively (from a $5D$ point of view, $\lambda$ is the tension of the brane). 

Thus the symmetric tensor $P_{\mu\nu}$ from STM can be interpreted as (proportional to) the total energy-momentum in a ${\bf Z}_2$ symmetric brane universe. This identification is also suggested  by the Hamiltonian treatment of five-dimensional Kaluza-Klein Gravity where $P_{\mu\nu}$ is the momentum conjugate to the induced metric $g_{\alpha\beta}$ \cite{Sajko1}. 

With this identification, (\ref{EMT in brane theory}), (\ref{conservation for T}) and (\ref{sources for Weyl tensor}) become the dynamical equations for gravity in the brane. These are identical to the ones developed in usual brane theory \cite{Maartens2}. 

Consequently, we conclude that the STM equations can be interpreted as the equations for gravity in a ${\bf Z}_2$ symmetric universe whose matter content is described by (\ref{P in terms of S}), with the usual decomposition (\ref{decomposing the brane matter tensor}), plus local and non-local corrections given by $\Pi_{\mu\nu}$ and $E_{\mu\nu}$, respectively. 
 In other words, the STM picture forms the generating space for
 brane-world solutions. 

The effective matter content of the spacetime will be the same whether we interpret it as induced matter, as in STM, or as the ``total" matter in a ${\bf Z}_2$ symmetric brane universe. Again, this is a consequence of the identification (\ref{P in terms of S}) and the field equation (\ref{EMT in brane theory}).

Since the dynamics of the spacetime is determined by its total matter content, we conclude that   these two theories generate the same physics, although STM and brane theories look very different at first sight. 
 
\subsection{Geometrization of Matter in Brane Models }

From (\ref{P in terms of the extrinsic curvature}) and (\ref{P in terms of S}) we see that the matter terms arise from the extrinsic curvature of the brane $\Sigma$. Equation (\ref{EMT in the brane}) then shows that the dependence of five-dimensional metrics on the extra coordinate leads to usual matter in $4D$. In the case where the $5D$ bulk metric is independent of the extra coordinate, $S_{\mu\nu} = 0$ and there is no matter on the brane. 

In this case, there is only an effective energy momentum induced on the brane by the Weyl curvature in the bulk projected onto the brane, viz., $^{(4)}G_{\mu\nu} = - \epsilon E_{\mu\nu}$. The induced matter is called Weyl (or dark) radiation, because it satisfies the ``radiation-like" equation of state $\rho = \sum_{i = 1}^{3} p_{i}$, where $\rho$ and $p_{i}$ are the energy density and principal pressures, respectively. The Davidson and Owen solution \cite{Davidson} is a perfect example of this and has been much discussed in STM theory .

Thus, although it is not mentioned in an explicit way, brane-world models incorporate the concept that  matter in $4D$ can be regarded as the effect of curvature in the extra dimension in a five-dimensional bulk. This is the typical point of view adopted in STM theory  \cite{Wesson book}. 

\subsection{Different Points of View in STM and Brane Theories}

We already mentioned that these two theories have different motivation for the introduction of a large extra dimension. Besides this, the main difference is that authors in STM and brane theory  adopt different points of view regarding the matter content and dynamics of our $4D$ spacetime. 

In STM theory the attention is on the geometry of the bulk. The bulk is a solution of the $5D$ Einstein's equations in vacuum. The matter content of the spacetime (brane) $\Sigma$ is  regarded as the effect of the curvature in the extra dimension. The geometry of the brane is determined by the $5D$ line element evaluated at $\Sigma$ defined as $y = y_{0}$. As an illustration of this procedure we mention the cosmological solution \cite{JPdeL 1}
\begin{equation}
\label{Ponce de Leon solution}
d{\cal S}^2 = y^2 dt^2 - t^{2/\alpha}y^{2/(1 - \alpha)}[dr^2 + r^2(d\theta^2 + \sin^2\theta d\phi^2)] - \alpha^2(1- \alpha)^{-2} t^2 dy^2,
\end{equation}
where $\alpha$ is a constant. In four-dimensions (on the hypersurface $\Sigma: y = y_{0}$) this metric corresponds to Friedmann-Robertson-Walker models with flat $3D$ sections. The equation of state of the effective perfect fluid in $4D$ is: $p = n \rho$ with $n = (2\alpha/3 -1)$ ($\alpha = 2$ for radiation, $\alpha = 3/2$ for dust, etc.).

In brane theory the main thrust is the matter content of the brane. When we assume some equation of state we impose conditions on the geometry of the bulk (specifically on $\stackrel{\ast}{g}_{\mu\nu}$). The ``backreaction" from the bulk shows up in the gravity on the brane through the nonlocal Weyl anisotropic  radiation field described by the trace-free tensor $E_{\mu\nu}$ (which contains $\stackrel{\ast \ast}{g}_{\mu\nu}$). Therefore, the equations in the brane, for some specific $T_{\mu\nu}$, should be solved together with the equations in the bulk, which are assumed to be the $5D$ Einstein's equation with negative cosmological constant. 

In short we can say that authors in STM construct $4D$ from $5D$, while authors in brane-world scenarios seek to reconstruct $5D$ from $4D$.

\section{Combining The Two Theories}

In this Section we approach the question of how can we benefit from the equivalence between these two theories.

In STM theory, the solutions of the field equations contain certain number of arbitrary functions. The interpretation of these solutions in terms of induced matter, as well as the physical properties, crucially depends on the choice of these functions. However, the number of physical restrictions, in the theory, are not in general sufficient to determine all of them.

In brane theory, the system of equations (\ref{EMT in brane theory}), (\ref{conservation for T}) and (\ref{sources for Weyl tensor}) is not in general closed, since (\ref{sources for Weyl tensor}) does not determine $E_{\mu\nu}$, in general. If $E_{\mu\nu} = 0$, then the field equations on the brane form a closed system. However, there is no guarantee that the resulting brane metric can be embedded in a regular bulk. Simply there is no enough information available in the brane as to reconstruct the bulk \cite{Cristiano}. 

Here we propose an alternative method to alleviate the task of finding solutions in the brane. Our proposal combines both theories. The idea is to capitalize from the rich physics in brane models and the freedom in STM.  Specifically,  we can impose the physics on the brane to restrict the freedom characteristic of many solutions in STM \cite{Liu Wesson}
-\cite{Fukui Seahra and Wesson}. 

\subsection{An Exact Solution in STM}

As an illustration of our proposal, let us consider the STM solution \cite{Liu Wesson}
\begin{equation}
\label{STM solution}
d{\cal S}^2 = B^2(t,y)dt^2 - A^2(t,y)\left[ \frac{dr^2}{1 - kr^2} + r^2 d\Omega^2\right] - dy^2,
\end{equation}
with
\begin{equation}
B = \frac{1}{\mu(t)}\frac{\partial A}{\partial t},
\end{equation}
and
\begin{equation}
\label{metric coeff. A}
A^2 = [\mu^2(t) + k]y^2 + 2 \nu y + \frac{\nu^2(t) + K}{\mu^2(t) + k}.
\end{equation} 
Here the extra dimension is spacelike, $\epsilon = -1$. This is an exact cosmological solution, with curvature index $k = -1, 0, +1$, which contains two arbitrary functions $\mu(t)$ and $\nu(t)$. The constant $K$ is related to the Kretschmann scalar, namely,
\begin{equation}
I = R_{ABCD}R^{ABCD} = \frac{72 K^2}{A^8}
\end{equation}
The effective total energy density and isotropic pressure are
\begin{equation}
\label{effective density STM}
8\pi G_{N}\rho_{eff} = ^{(4)}G^{0}_{0} =  \frac{3(\mu^2 + k)}{A^2},
\end{equation}
\begin{equation}
8 \pi G_{N}p_{eff} = - ^{(4)}G^{1}_{1} = - ^{(4)}G^{2}_{2} = - ^{(4)}G^{3}_{3} = - \frac{2\mu \dot{\mu}}{A\dot{A}} - \frac{\mu^2 + k}{A^2}.
\end{equation}
This solution has been studied in detail \cite{Liu Wesson} for various choices of $\mu(t)$ and $\nu(t)$ and imposing the equation of state $p_{eff} = n\rho_{eff}$, with $n = Constant$. Certainly, this physical assumption decreases the number of unknown functions from two to one. But still there is a lot of freedom. 
\subsection{An Exact Solution on The Brane}
Here we will use the brane-world paradigm to determine the unknown  functions. The usual assumption is that normal matter, except for gravity, ``lives" only on the $\Sigma: y = 0$ brane \cite{Shiromizu}-\cite{Cristiano}. The appropriate ${\bf Z}_2$ symmetric brane universe, from the generating space (\ref{STM solution})-(\ref{metric coeff. A}), is obtained following (\ref{Z2 symmetry}).

The junction conditions across the brane imply that the metric is continuous. Specifically, on $\Sigma$
\begin{equation}
\label{FRW from the brane}
ds^2 = dt^2 - a^2(t)\left[ \frac{dr^2}{1 - kr^2} + r^2 d\Omega^2\right],
\end{equation}
where 
\begin{equation}
A^2(t,0) = a^2(t) = \frac{\nu^2 + K}{\mu^2 + k},
\end{equation}
\begin{equation}
B(t, 0) = \left(\frac{\dot{A}}{\mu}\right)_{|\Sigma} = \frac{\dot{a}}{\mu} = 1.
\end{equation}
Thus the continuity of the metric defines $\mu$ and $\nu$ through $a$,
\begin{eqnarray}
\label{mu and nu in terms of a}
\mu &=& \dot{a}\nonumber  \\
\nu^2 &=& a^2({\dot{a}}^2 + k) - K.
\end{eqnarray}
We now use the identification of $P_{\mu\nu}$ with the energy-momentum tensor on the brane. From equations (\ref{EMT in the brane}), (\ref{P in terms of S}) and (\ref{decomposing the brane matter tensor}) we get 

\begin{equation}
\label{density eq.}
- \lambda + \rho = 3\alpha \left(\frac{\stackrel{\ast}{A}}{A}\right)_{|\Sigma} = 3 \frac{\nu\alpha}{a^2},
\end{equation}
\begin{equation}
\label{pressure eq.}
- \lambda - p = \alpha \left(\frac{\stackrel{\ast}{B}}{B} + 2 \frac{\stackrel{\ast}{A}}{A}\right)_{|\Sigma} = \frac{\dot{\nu}\alpha}{a\dot{a}} + \frac{\nu \alpha}{a^2},
\end{equation}
where $\alpha = 2/k_{(5)}^2$. Here we have assumed that the energy-momentum tensor $T_{\mu\nu}$ in the brane is a perfect fluid with density $\rho$ and isotropic pressure $p$. This fluid is ``at rest" in the frame of  (\ref{FRW from the brane}) because $^{(4)}G_{0j} = 0$. In other words, the system of reference in (\ref{FRW from the brane})
is comoving with the matter. 

As in brane models we impose some physics on the matter quantities $\rho$ and $p$ (not on the effective ones). We will adopt the usual equation of state in cosmology 
\begin{equation}
p = n\rho,
\end{equation}
which for $n = 1, 3, 0$ gives the stiff, radiation and dust equations of state. Substituting into (\ref{density eq.}) and (\ref{pressure eq.}) we get
\begin{equation}
\frac{d\nu}{da} + (3n + 1)\frac{\nu}{a} = - \frac{\lambda(n + 1)}{\alpha}a.
\end{equation}
From which we find
\begin{equation}
\label{nu in terms of a}
\nu = - \frac{ \lambda}{3\alpha} a^2 + \frac{C}{a^{(1 + 3n)}},
\end{equation}
where $C$ is a constant of integration. The matter energy density becomes
\begin{equation}
\label{matter density}
\rho = \frac{ 3\alpha C}{a^{3(1 + n)}}. 
\end{equation} 
The evolution equation for $a$ is obtained from (\ref{mu and nu in terms of a}) and(\ref{nu in terms of a}) as 
\begin{equation}
\label{evolution of a}
{\dot{a}}^2
= - k + \frac{K}{a^2} + \frac{\lambda^2}{9\alpha^2}a^2 + \frac{C^2}{a^{2(2 + 3n)}} - \frac{2 C \lambda}{3 \alpha a^{(1 + 3n)}}.
\end{equation}
For a perfect fluid, $T_{\mu\nu} = (\rho + p)u_{\mu}u_{\nu} - pg_{\mu\nu}$, the quadratic corrections in the brane (\ref{quadratic corrections}) become
\begin{equation}
\Pi_{\mu\nu} = - \frac{\rho^2}{12}u_{\mu}u_{\nu} - \frac{1}{12}\rho(\rho + 2p)h_{\mu\nu},
\end{equation}
where we have introduced the projector $h_{\mu\nu} = u_{\mu}u_{\nu} - g_{\mu\nu}$. Thus, for perfect fluid $\Pi^{0}_{0} = - \rho^2/12$. Therefore, the quadratic contribution to the total energy density is always  positive, which is physically reasonable (we remind that $\epsilon = -1$ here). The quadratic correction to the pressure is also positive and proportional to $(2n + 1)\rho^2/12$.

The black or Weyl radiation coming from the bulk is described by $E_{\mu\nu}$. It does not depend on the equation of state in the brane. The corresponding energy density is
\begin{equation}
\label{Weyl radiation}
8 \pi G _{N}\rho_{Weyl} = - \epsilon E_{0}^{0} = - \frac{\stackrel{\ast \ast}{B}}{B} = \frac{3 K}{a^4},
\end{equation}
where the factor $8 \pi G_{N}$ is introduced for dimensional reasons.
The radiation pressure is isotropic, viz., $8\pi G_{N}p_{Weyl} = - E_{1}^{1} =  - E_{2}^{2} = - E_{3}^{3} = K/a^4$, and $E_{\alpha}^{\alpha} = 0$, as expected.

Collecting results, we find the effective energy density as follows
\begin{equation}
\label{General effective density}
8 \pi G_{N} \rho_{eff} = \Lambda + 8 \pi G_{N}\rho + \frac{k_{(5)}^4}{12}\rho^2
+ 8\pi G_{N}\rho_{Weyl},
\end{equation}
where $\Lambda = \lambda^2k_{(5)}^2/12$; and  the densities $\rho$, $\rho_{Weyl}$ are given by (\ref{matter density}) and (\ref{Weyl radiation}), respectively. We should remark again that the effective matter content is the same whether calculated from STM equations or from the ${\bf Z_2}$-symmetric brane perspective. This is verified by our example, viz., the STM equation (\ref{effective density STM}) and the brane-world equation (\ref{General effective density}) yield the same expression for $\rho_{eff}$, in terms of $a$. We also note that (\ref{General effective density}) is a general expression valid for any perfect fluid and not only for this model. Another important point to notice here is that the Weyl contribution to the effective energy density does not have to be positive. It can be positive, negative or zero \cite{Old JPdeL}.

For very high energies in the brane, when the quadratic term dominates over the other terms, the  effective  equation of state is stiffened, viz., 
\begin{equation}
p_{eff} \approx (2n + 1)\rho_{eff}. 
\end{equation} 
This is a general feature of brane models \cite{Maartens2}. It indicates that quadratic corrections might have a dramatic influence in the early universe and during late-stages of the gravitational collapse. Even in the case of ``non-gravitating" matter, for which $(\rho + 3p) = 0$, the effective matter is radiation-like, viz., $p_{eff} \approx \rho_{eff}/3$. For dust $p_{eff} \approx \rho_{eff}$. This feature presents a number of intriguing possibilities that is worth to study in detail.

The above equations  represent a cosmological model where the brane  is appropriately matched to the bulk metric and the matter in the brane satisfies physical conditions. This model has a number of interesting properties but we leave their discussion to another opportunity.

This model is an example that clearly illustrates the feasibility of our proposal here. Namely, that a combination of the two approaches, STM and brane theory, can be very helpful to find solutions on the brane. This is a working alternative to the direct approach in brane theory of solving the equations from  ``scratch".

\section{Summary and Conclusions}

We have presented here a new interpretation of space-time-matter theory. In this interpretation the equations of  STM are identical to those in ${\bf Z}_2$-symmetric brane-world models. This is a consequence of the connection between the symmetric tensor $P_{\alpha\beta}$ from STM and the energy-momentum tensor $S_{\mu\nu}$ in the brane. This, in turn is a result of the ${\bf Z}_2$ symmetry used in brane-wold theory. Therefore, STM incorporates all the local and non-local energy corrections to GR in $4D$, typical of  brane-world scenarios. 

In both theories the presence of matter crucially relies on the assumption that the metric is a function of the extra variable. For metrics with no dependence on $y$ there is no matter, just Weyl radiation with anisotropic pressures. In this sense, matter in $4D$ is a consequence of the extra dimension. Thus, brane models incorporate the concept that matter can be regarded as effect of the geometry in $5D$, which is the trademark  of STM theory. 

Going a little aside, we mention that the motion of test particles presents similar characteristics in both theories. This has been studied in a number of papers \cite{Wesson Mashhonn}-\cite{last work}. However, this similarity has nothing to do with the dynamics of the field, but it is a result of the assumption that test particles move along geodesics in both theories. 

From a theoretical point of view it is important that we understand the connection between theories that seek to explain the same subject. STM and brane models represent two opposite approaches to the same problem. In STM the physics in $4D$ is constructed from the bulk. In brane models we should find the solution to the equations in $4D$, which has to be matched with the bulk geometry satisfying the appropriate boundary conditions. In other words, the goal in brane theory is to use physical information in $4D$ to reconstruct the generating bulk.  

From a practical point of view, a combined approach of both theories seems to be promising. The incorporation of the rich physics of branes to STM may allow us to obtain interesting physical models. The solution discussed  in Section 3 is a clear example of this. It neatly shows how the physics on the brane unambiguously determines the arbitrary functions. It also serves as an example of our main point here. Namely, that effective matter quantities do not depend on whether we calculate them using the STM or brane-world paradigm; only the interpretation is different in both theories. This example-solution presents good physical properties, but their discussion is beyond the scope of the present work.

The main difference between the two models resides in the motivation they have for the introduction of a large extra dimension. Besides the bulk geometry in STM satisfies $^{(5)}G_{AB} = 0$, while in brane theory $^{(5)}G_{AB} = - \Lambda_{(5)} \gamma_{AB}$. For this reason there is a ``missing" term in the definition of $\Lambda_{(4)}$ in (\ref{definition of lambda}). The introduction of this term presents no problem and one would get
\begin{equation}
\Lambda_{(4)} = \frac{1}{2}k_{(5)}^2\left(\Lambda_{(5)} + \epsilon\frac{k_{(5)}^2\lambda^2}{6}\right).
\end{equation}

Despite of these differences, we conclude that STM and ${\bf Z}_2$-symmetric brane-world theories are equivalent to each other. They predict the same corrections to general relativity in $4D$, and incorporate the concept that matter can be regarded as the effect from an extra dimension.

\end{document}